# Fluorides of silver under large compression

Dominik Kurzydłowski,[a*] Mariana Derzsi,[b,c] Eva Zurek,[d] Wojciech Grochala[c]

[a.] *Faculty of Mathematics and Natural Sciences, Cardinal Stefan Wyszyński University, Warsaw 01-038, Poland*

[b.] *Advanced Technologies Research Institute, Faculty of Materials Science and Technology in Trnava, Slovak University of Technology in Bratislava, Jána Bottu 8857/25, 917 24 Trnava, Slovakia*

[c.] *Centre of New Technologies, University of Warsaw, ul. Banacha 2c, Warsaw 02-097, Poland*

[d.] *Department of Chemistry, State University of New York at Buffalo, New York 14260-3000, United States*

The silver-fluorine phase diagram has been scrutinized as a function of external pressure using theoretical methods. Our results indicate that two novel stoichiometries containing $Ag^+$ and $Ag^{2+}$ cations ($Ag_3F_4$ and $Ag_2F_3$) are thermodynamically stable at ambient and low pressure. Both are computed to be magnetic semiconductors at ambient pressure conditions. For $Ag_2F_5$, containing both $Ag^{2+}$ and $Ag^{3+}$, we find that strong 1D antiferromagnetic coupling is retained throughout the pressure-induced phase transition sequence up to 65 GPa. Our calculations show that throughout the entire pressure range of their stability the mixed valence fluorides preserve a finite band gap at the Fermi level. We also confirm the possibility of synthesizing $AgF_4$ as a paramagnetic compound at high pressure. Our results indicate that this compound is metallic in its thermodynamic stability region. Finally, we present general considerations on the thermodynamic stability of mixed valence compounds of silver at high pressure.

## Introduction

The reactivity of elemental fluorine at pressures exceeding 1 GPa (= 10 kbar) has recently attracted increased scientific attention because of the exotic nature of materials that are predicted to form in fluorine-rich systems at large compression.[1–14] These include compounds featuring elements in high oxidation states (e.g. $IF_8$),[13] hypervalent second-row elements ($NF_5$),[8] as well as transient species that are not attainable at ambient conditions ($AuF_2$,[3,4] $AuF_6$). In the case of transition metal fluorides, this novel chemistry might lead to species featuring exotic magnetic ions, such as $Au^{4+}$ (*5d⁷* electronic configuration) or $Au^{2+}$ (*5d⁹*). This provides a link between the high-pressure chemistry of $M/F_2$ systems (M – transition metal) and investigations concerned with the influence of compression on the magnetic properties of materials.[15–17]

In this context, the study of the $Ag/F_2$ system under large compression is of particular interest. Recent calculations predict that $AgF_4$, containing the $Ag^{4+}$ cation (*4d⁷* electronic configuration), should become thermodynamically stable at high pressures.[3] On the other hand, compression-induced phase transitions of silver difluoride, $AgF_2$ (containing the *4d⁹* $Ag^{2+}$ cation),[18,19] are found to lead to both a dramatic increase in the antiferromagnetic (AFM) superexchange interaction between $Ag^{2+}$ centres, and a decrease of the magnetic dimensionality from 2D to 0D.[20] The interest in spin-spin interactions in this system is motivated by recent discoveries of many similarities between $AgF_2$ and $La_2CuO_4$ – a precursor of copper oxide-based superconductors.[21–24]

Here we present a detailed computational exploration of the phase diagram of the binary $Ag/F_2$ system under large compression. In contrast to recent work,[3] which considered only compounds with Ag in a single valence state ($Ag^IF$, $Ag^{II}F_2$, $Ag^{III}F_3$, $Ag^{IV}F_4$), we also looked at the possibility of formation of as yet unknown mixed-valence compounds such as $Ag_3F_4$ (= $Ag^I_2Ag^{II}F_4$) or $Ag_2F_5$ (= $Ag^{II}FAg^{III}F_4$). The mixed-valence systems are of particular interest here since they formally correspond to electron- or hole-doped $AgF_2$. Moreover, either a substantial decrease or even closing of the fundamental band gap (metallization) is expected to be obtained upon $e^-$ or $h^+$ doping in mixed-valence or intermediate valence fluorides of silver, as typical examples of Class I or Class III mixed-valence compounds,[25] respectively.

## Computational details

**DFT calculations:** Periodic DFT calculations of the geometry, enthalpy, and electronic structure of various $Ag_mF_n$ phases up to 100 GPa utilized the SCAN meta-GGA functional.[26] This functional was found to offer an accurate description of the high-pressure properties of fluorides of late transition metals (zinc,[27] gold,[28] and silver[20]). In addition, calculations utilizing SCAN yielded geometric parameters of the ambient pressure structure of $AgF_2$ that were in line with experiment,[29] with differences in lattice parameters and bond lengths not exceeding 1.5 %. This functional also performs well in terms of thermodynamic properties, such as atomization energies.[30] We find that it reproduces both the enthalpy of formation of AgF (theory: –210 kJ/mol, experiment: –206 kJ/mol), and the dissociation energy ($D_0$) of $F_2$ (theory: 1.62 eV, experiment: 1.63 eV[31]).

The thermodynamic stability of various structures within a given $Ag_mF_n$ system was judged by comparing their enthalpy ($H$), and thus the calculations formally correspond to $T$ = 0 K at which the Gibbs free energy ($G = H - S \cdot T$, where $S$ is the entropy) is equal to $H$. We did not include the zero-point energy correction in the calculations.

The thermodynamic stability of various $Ag_mF_n$ compounds was judged by comparing their enthalpies of formation, $H_f(Ag_mF_n)$ (in eV per atom), which at a given pressure are calculated from the following expression:

$$H_f(Ag_mF_n) = \frac{H(Ag_mF_n) - mH(Ag) - n/2H(F_2)}{m+n}$$

Where $H(Ag_mF_n)$ is the enthalpy of the most stable $Ag_mF_n$ structure at a given pressure, while $H(Ag)$ is the enthalpy of gold in the *fcc* ($Fm\overline{3}m$) structure,[32] and $H(F_2)$ is the enthalpy of the molecular crystal of the $F_2$ (α polymorph).[8,33]

The projector-augmented-wave (PAW) method,[34] as implemented in the VASP 5.4 code,[35,36] was used in the calculations. The cut-off energy of the plane-waves was set to 800 eV with a self-consistent-field convergence criterion of $10^{-6}$ eV. The valence electrons (Ag: $4d^{10}$, $5s^1$; F: $2s^2$, $2p^5$) were treated explicitly, while standard VASP pseudopotentials, accounting for scalar relativistic effects, were used for the description of core electrons. The k-point mesh spacing was set to $2\pi \times 0.03$ Å$^{-1}$. All structures were optimized until the forces acting on the atoms were smaller than 0.015 eV/Å. To account for the open-shell nature of $Ag^{2+}$ cations (*$4d^9$* electronic configuration), all calculations were performed with spin-polarization. For $Ag^{2+}$-bearing compounds, the geometry optimizations were performed for the lowest-energy spin state. The electronic density of states (DOS) was calculated both using the SCAN method, as well as the HSE06 hybrid functional,[37] in both cases SCAN-optimized structures were used.

Values of the superexchange coupling constants (defined by the Heisenberg Hamiltonian in the form: $\boldsymbol{H}_{ij} = -J_{ij}\cdot\boldsymbol{s}_i\cdot\boldsymbol{s}_j$) between $Ag^{II}$ sites in the $P\overline{1}(β_1)$, $P\overline{1}(β_2)$, and $P\overline{1}(β_3)$ structures of $Ag_2F_5$ (in the 0 – 30 GPa pressure range) were extracted via the broken-symmetry method,[38,39] as applied in refs.[20,40] The applied definition of the Heisenberg Hamiltonian leads to positive values of $J_{ij}$ for ferromagnetic (FM) coupling, and negative for an antiferromagnetic (AFM) one.

Evolutionary algorithm searches were performed to identify the lowest-enthalpy structures of $Ag_mF_n$ compounds at ambient and high pressure. Searches were performed for seven stoichiometries: $Ag_2F$, $Ag_3F_4$, $Ag_2F_3$, $Ag_3F_5$, $Ag_2F_5$, $Ag_3F_8$, $AgF_3$, $AgF_4$. For this we used the XtalOpt software (version r12)[41,42] coupled with spin-polarized DFT calculations utilizing the PBE functional.[43] These searches were conducted at 10, 40, and 80 GPa for Z equal to 1, 2, 3, 4 and 6.

Visualization of all structures was performed with the VESTA software package.[44] For symmetry recognition we used the FINDSYM program.[45]

## Results and discussion

Different compounds in the Ag-F binary system can be grouped according to their fluorine content ($x_F$, which we here give in mol %), and the resulting oxidation state of the silver atom. The three simplest silver fluorides stable at ambient conditions, AgF ($x_F$ = 50 mol %), $AgF_2$ ($x_F$ = 66 mol %), and $AgF_3$ ($x_F$ = 75 mol %) – all of them insulators – contain silver in the +I, +II, and +III oxidation state, respectively. Compounds with a fluorine content smaller than that of AgF should contain silver in a formal oxidation state lower than +I. This is the case of metallic silver sub-fluoride, $Ag_2F$ ($x_F$ = 33 mol %),[46] for which the silver atom can be assigned a formal oxidation state of +½. For fluorine contents intermediate between that of AgF and $AgF_2$, one should expect the formation of mixed-valent compounds containing both $Ag^+$ and $Ag^{2+}$ cations, such phases however have not been synthesized to date, although their existence has been speculated.[47] In contrast, two mixed-valent fluorides containing $Ag^{2+}$ and $Ag^{3+}$ cations, with fluorine content intermediate between that of $AgF_2$ and $AgF_3$, are known. These are: $Ag_2F_5$ ($x_F$ = 71 mol %),[48–50] which can be written as $Ag^{(II)}F[Ag^{(III)}F_4]$, and $Ag_3F_8$ ($x_F$ = 73 mol %),[48,51] which is better formulated as $Ag^{(II)}[Ag^{(III)}F_4]_2$. Finally, one might envisage a silver tetrafluoride, $AgF_4$ ($x_F$ = 80 mol %) containing silver in the exotic +IV oxidation state. This compound has not been synthesized to date, although it was recently predicted to be thermodynamically stable above 37 GPa.[3]

To assess the thermodynamic stability of different silver fluorides at ambient and high pressure, we used the enthalpies of their most stable phases, obtained from DFT calculations (see Computational Methods), to construct the so-called 'convex hull' diagram – a plot of the formation enthalpy ($H_f$) (in eV per atom) as a function of F-content. Stoichiometries lying on the convex hull are thermodynamically stable, while those lying above it are unstable with respect to decomposition into stable compounds neighbouring them on the convex hull.

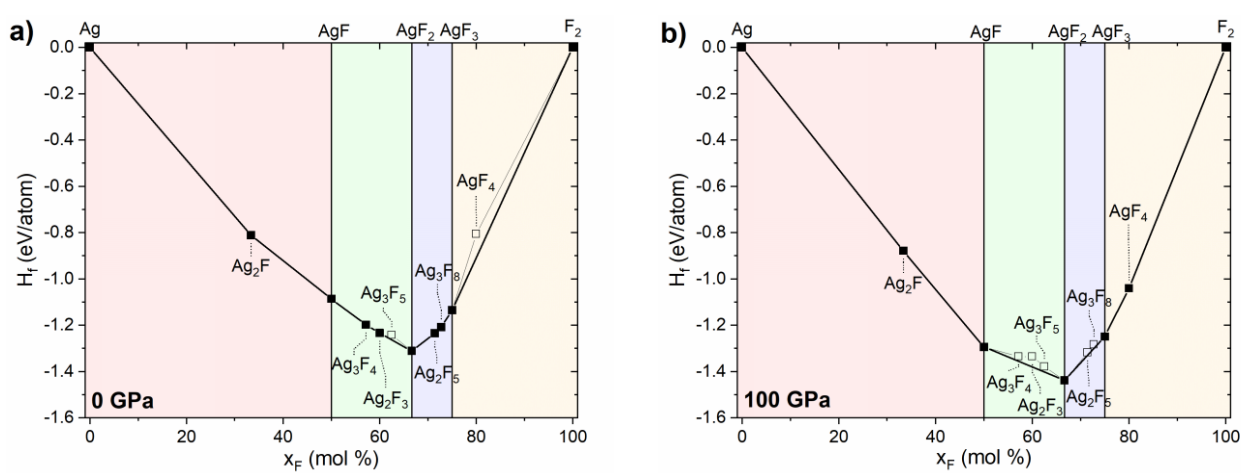

**Figure 1** Calculated convex hull diagrams for the Ag/F system at 0 (a) and 100 GPa (b). Stoichiometries forming the convex hull are marked with filled squares, those above it with empty squares. Colours indicate regions of fluorine content corresponding to different oxidation states of Ag atoms (red – below +1, green – mixed-valent +1/+2, blue – mixed valent +2/+3, orange – above +3).

As can be seen in Figure 1a, our DFT calculations reproduce the thermodynamic stability of $Ag_2F$, AgF, $AgF_2$, $Ag_2F_5$, $Ag_3F_8$, and $AgF_3$ at ambient pressure (effectively 0 GPa). It is noteworthy that although thermodynamically stable with respect to other Ag/F binaries, fluorides containing $Ag^{2+}$ and $Ag^{3+}$ cations ($AgF_2$, $Ag_2F_5$, $Ag_3F_8$, and $AgF_3$) are very reactive and moisture-sensitive. Moreover, $AgF_3$ is thermally fragile, and when dissolved in anhydrous HF tends to partly release $F_2$ and transform into $Ag_3F_8$ at room temperature.[48] Interestingly, we find two mixed valent $Ag^+/Ag^{2+}$ fluorides that are thermodynamically stable at 1 atm – $Ag_3F_4$ (= $Ag^{(I)}{}_2Ag^{(II)}F_4$) and $Ag_2F_3$ (= $Ag^{(I)}Ag^{(II)}F_3$). We will describe these in more detail further in the text.

The effect of compression on the phase stability of the Ag/F system is seen best when comparing the convex hull at 0 GPa (Figure 1a) with that obtained at 100 GPa (Figure 1b) – the limiting pressure of this study. Compression leads to destabilization of the mixed-valent fluorides. Those containing $Ag^+/Ag^{2+}$ cations become thermodynamically unstable at 0.9 and 20 GPa for $Ag_2F_3$ and $Ag_3F_4$, respectively (Figure 2). The $Ag^{II}F[Ag^{III}F_4]$ stoichiometry persists up to 83 GPa, while $Ag^{II}[Ag^{III}F_4]_2$ is stable up to 18 GPa. However, even at 100 GPa mixed-valent fluorides lie close to the convex hull (not more than 50 meV/atom above it), suggesting that these compounds could be metastable at high pressure.[52] On the other hand, high pressure leads to stabilization of $AgF_4$ (at 56 GPa), in accordance with a recent study;[3] here, the formation pressure is predicted to be ca. 20 GPa higher than the previously reported one, which largely stems from the different DFT functionals used in the calculations.

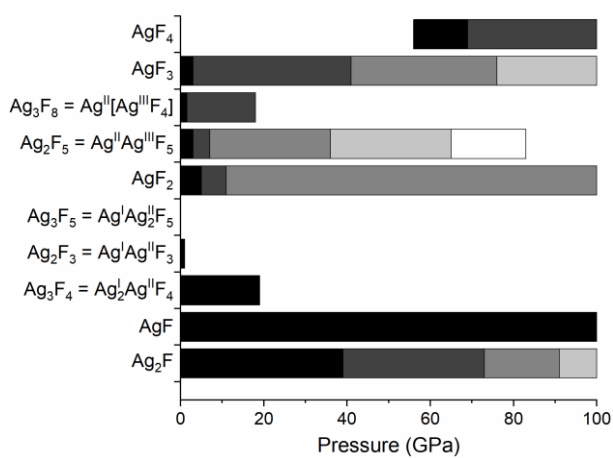

**Figure 2** Bar diagram showing the calculated thermodynamic stability of different silver fluoride stoichiometries as a function of pressure, up to 100 GPa. Stability regions of different polymorphs are indicated.

As can be seen in Figure 2, silver fluorides exhibit a rich polymorphism when squeezed. We now move on to the description of the pressure-induced structural changes in each of the $Ag_mF_n$ compounds within their thermodynamic stability range. We start with two fluorides that were studied experimentally at high pressure: AgF and $AgF_2$. For the latter compound diamond anvil cell (DAC) experiments indicated that the distorted HP-$PdF_2$-type structure (space group *Pbca*, $Z$ = 4) transforms to a structure with $Pca2_1$ ($Z = 4$) symmetry at 10 GPa, which in turn transforms to a distorted HS-ZrO$_2$-type[27] phase ($Pbcn$ symmetry, $Z = 8$) at 15 GPa.[19] Our calculations reproduce this phase sequence, with the phase transition pressures underestimated by 4 GPa ($Pbca \xrightarrow{6\,GPa} Pca2_1 \xrightarrow{11\,GPa} Pbcn$). For AgF, the CsCl-type structure (space group $Pm\bar{3}m$, $Z = 1$) is calculated to be the most stable phase within the whole pressure range studied (0 – 100 GPa). At ambient conditions AgF crystallizes in the NaCl-type structure ($Fm\bar{3}m$ symmetry, $Z = 4$), however, a phase transition to the CsCl structure type is observed already at 2.7 GPa.[53] It seems that the extended stability range of this phase in our calculations is a result of a slight underestimation of the transition pressure between the NaCl and CsCl structures.

**Silver sub-fluoride (Ag$_2$F)**

At ambient conditions, Ag$_2$F adopts the layered anti-CdI$_2$ structure ($P\bar{3}m$, $Z = 1$) with six-fold coordination of F, and three-fold of Ag (Figure 3a – for cif files of selected structures see the Electronic Supplementary Information).[54] This structure consists of Ag$_2$F layers with an AA-type stacking. Calculations indicate that this structure should transform at 39 GPa into an anti-CdCl$_2$ polymorph ($R\bar{3}m$, $Z = 3$) that differs from $P\bar{3}m$ by a stacking of the layers (AB-type). We predict that further compression should induce a transition into a tetragonal structure of $I4/mmm$ symmetry ($Z = 2$) in which the coordination number of F and Ag increases to 8 and 4, respectively (Figure 3b). Finally, above 91 GPa the $P4/mmm$ phase ($Z = 2$) becomes the lowest-enthalpy structure of Ag$_2$F. This polymorph can be described as an intergrowth of *fcc* Ag (a slab of 3 atomic layers) with CsCl-type AgF, with the two slabs sharing a common Ag layer (Figure 3c). The Ag-Ag distances within the Ag slab (2.57 Å at 100 GPa) are almost identical to the nearest-neighbour Ag-Ag distance calculated for the *fcc* structure of silver at the same pressure (2.58 Å). The same holds when comparing the Ag-F distances within the AgF slab (2.28 Å) with those calculated for the CsCl polymorph of AgF (2.27 Å).

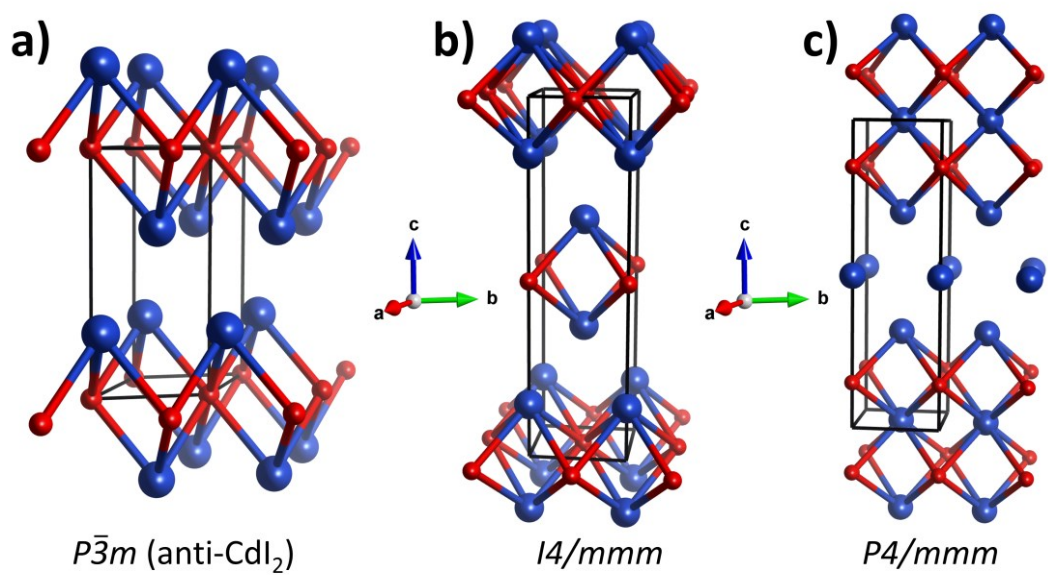

**Figure 3** Selected low-enthalpy Ag$_2$F phases: $P\bar{3}m$ (a), $I4/mmm$ (b), $P4/mmm$ (c). Blue/red balls depict Ag/F atoms.

The appearance of the *P4/mmm* structure might signal the tendency of $Ag_2F$ to disproportionate into AgF and Ag at pressures above 100 GPa. Indeed, the enthalpy of reaction $Ag_2F \rightarrow Ag + AgF$ decreases from 25.2 kJ/mol at ambient pressure to just 4.4 kJ/mol at 100 GPa. However, up to 100 GPa $Ag_2F$ remains on the convex hull of the Ag/F binary phase diagram.

The phase transition sequence predicted for $Ag_2F$ can be compared to that observed for $Ca_2N$.[55,56] At ambient pressure $Ca_2N$ adopts a structure resembling that of $Ag_2F$, but $Ca_2N$ is, however, a 2D electride system with excess electrons localized between $[Ca_2N]^+$ layers.[57] As a result the phase transition sequence is quite different in the two system with $Ca_2N$ evolving into structures with reduced dimensionality (1D to 0D),[56] while $Ag_2F$ retains its 2D character. We did not find evidence of the emergence of electride behaviour of $Ag_2F$ upon compression.

### $Ag^+/Ag^{2+}$ mixed-valent compounds ($Ag_3F_4$, $Ag_2F_3$, and $Ag_3F_5$)

Before we move on to the description of the $Ag^I/Ag^{II}$ and $Ag^{II}/Ag^{III}$ mixed-valent fluorides we would like to note that depending on its oxidation state silver exhibits marked differences in its magnetism, and coordination by fluorine. Silver in the +I oxidation state is usually surrounded by at least 8 fluorine atoms with the shortest Ag-F distance larger than 2.25 Å. Additionally, $Ag^I$ sites do not host unpaired spin density, in line with the diamagnetic nature of the $Ag^+$ cation (*4d¹⁰* configuration). Silver in the +III oxidation state (*4d⁸*) is also diamagnetic, as it is always found in a square-planar coordination environment with Ag-F bonds shorter than 2 Å. This leaves $Ag^{II}$ (*4d⁹*) as the only magnetic silver species in the +I/+II and +II/+III mixed-valent formulations. Most often it is found either in an elongated octahedral geometry, with four short Ag-F distances (2.0 to 2.2 Å) forming a square, and two longer ones (longer than 2.3 Å) in apical positions, or in a compressed octahedral geometry with two short *trans* Ag-F distances (2.0 – 2.1 Å) and four longer ones (> 2.3 Å).

### $Ag_3F_4$

Our calculations indicate that at ambient conditions the most stable phase of $Ag_3F_4$ adopts a distorted spinel structure of the $CdMn_2O_4$-type ($I\bar{4}2d$ space group, $Z = 4$), rather than the post-perovskite structure proposed earlier.[47] This polymorph can be described as built from distorted $[Ag^{II}F_4]^{2-}$ squares separated by $Ag^+$ cations (Figure 4a). This interpretation is confirmed by: (i) the presence of unpaired electron density on the $[AgF_4]^{2-}$ units (magnetic moment equal to 0.84 Bohr magnetons, $\mu_B$); (ii) a short Ag-F bond within these units (2.13 Å at 10 GPa); (iii) lack of magnetization on the $Ag^+$ sites; and (iv) the large coordination number (CN = 8) and long Ag-F contacts (shortest one equal to 2.34 Å at 10 GPa) for the nonmagnetic Ag sites. This proves that the stoichiometry of $I\bar{4}2d$ is best described as $Ag^I_2Ag^{II}F_4$. This polymorph remains the ground state structure of $Ag_3F_4$ within its thermodynamic stability range (0 to 19 GPa); above this pressure $Ag_3F_4$ is predicted to decompose into AgF and $AgF_2$. The computed stability of $Ag_3F_4$ at ambient pressure conditions might explain the observation of photochemical decomposition of $AgF_2$ with the formation of species that yield a Raman spectrum typical to that observed for $M_2AgF_4$ salts.[58]

The $[AgF_4]^{2-}$ quasi-square planar plaquettes in the $I\bar{4}2d$ structure are distorted towards a tetrahedron with pairs of F atoms in a *trans* position shifted up/down by 0.3 Å with respect to the plane of the square. The next-nearest-neighbour contacts also form a tetrahedron as shown in Figure 4a. This can be compared with the coordination of $Cu^{2+}$ in the isostructural $CuCr_2O_4$ oxide.[59] In this case the departure of O atoms from the plaquette plane is much larger (0.9 Å), and the coordination polyhedron is closer to a tetrahedron.

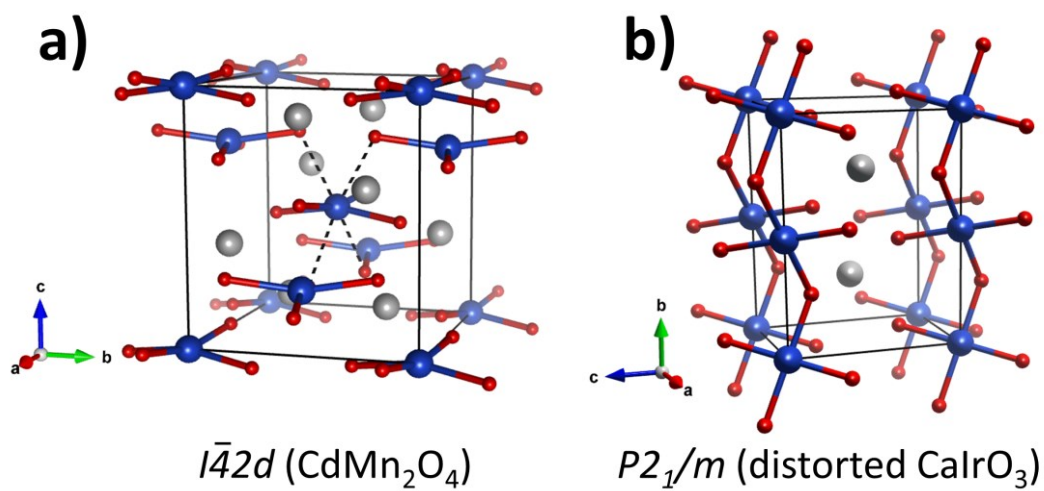

**Figure 4** The most stable phases of $Ag_3F_4$ (a) and $Ag_2F_3$ (b). Grey/blue/red balls mark $Ag^I$/$Ag^{II}$/F atoms. Dotted lines in (a) mark the second coordination sphere of $Ag^{II}$.

Given that the ionic radius of $Ag^+$ falls in between that of $Na^+$ and $K^+$,[60] one might expect that $Ag^I_2Ag^{II}F_4$ would adopt a layered perovskite structure (as found for α-$K_2AgF_4$),[61] or a post-perovskite phase (as found for β-$K_2AgF_4$ and $Na_2AgF_4$).[62–64] However, at ambient pressure both structure types have an enthalpy around 100 meV per $Ag_3F_4$ higher than $I\bar{4}2d$. This difference increases with pressure (250 meV at 5 GPa) due to the larger volume of the perovskite-derived phases.

**Ag$_2$F$_3$ and Ag$_3$F$_5$**

At ambient conditions, the most stable structure of Ag$_2$F$_3$ is of *P2$_1$/m* symmetry ($Z = 2$), and it can be described as a distorted variant of the post-perovskite CaIrO$_3$ structure (*Cmcm*, $Z = 2$). Analysis of the Ag-F distances and the spin density indicates that this polymorph can be described as being built from [Ag$^{II}$F$_3$]$^-$ chains separated by Ag$^I$ atoms (Figure 4b), as customary for M$^I$AgF$_3$ salts.[65–67]

At ambient pressure Ag$_2$F$_3$ is marginally stable with respect to decomposition into Ag$_3$F$_4$ and AgF$_2$ (the energy of the reaction: 2Ag$_2$F$_3$ → Ag$_3$F$_4$ + AgF$_2$ is positive by only 1.7 kJ/mol), and it becomes thermodynamically unstable above 0.9 GPa. The thermodynamic instability of Ag$_2$F$_3$ (=Ag$^I$Ag$^{II}$F$_3$) under large compression is analogous with the pressure-induced decomposition of other ABX$_3$ compounds, such as NaMgF$_3$,[68,69] MgGeO$_3$,[68] and MgSiO$_3$.[70] At high pressure all of these systems adopt the CaIrO$_3$-type structure, which upon compression decomposes to: (i) BX$_2$ and A$_2$BX$_4$ (as for MgGeO$_3$); (ii) AX and AB$_2$X$_5$ (as found for NaMgF$_3$), or (iii) both A$_2$BX$_4$ and AB$_2$X$_5$ (as in MgSiO$_3$). In all of these cases, A$_2$BX$_4$ is found to adopt the CdMn$_2$O$_4$ structure type. Our calculations reveal that Ag$_2$F$_3$ follows the first scenario with subsequent decomposition of A$_2$BX$_4$ into AX and BX$_2$. We have verified that for the studied system the AB$_2$X$_5$ stoichiometry (Ag$_3$F$_5$ = Ag$^I$Ag$^{II}$$_2$F$_5$) is not thermodynamically stable within the investigated pressure range, as evident from Figure 1. Summarizing this section, Ag$_2$F$_3$ is worth targeting in experiment at ambient (p,T) conditions.

**Ag$^{2+}$/Ag$^{3+}$ mixed-valent compounds (Ag$_2$F$_5$, Ag$_3$F$_8$)**

**Ag$_2$F$_5$**

At ambient conditions Ag$_2$F$_5$ (= Ag$^{II}$F[Ag$^{III}$F$_4$]) crystallizes in a $P\bar{1}$ ($Z = 4$) structure featuring chains of *trans*-connected [Ag$^{II}$F$_4$]$^{2-}$ plaquettes interconnected by [Ag$^{III}$F$_4$]$^-$ squares, as found by x-ray diffraction measurements.[50] Our calculations confirm that this structure, which we will refer to as $P\bar{1}(\alpha)$, has the lowest energy at 1 atm. In accordance with the oxidation state assignment, we find [Ag$^{II}$F$_4$]$^{2-}$ units to host unpaired electron density, and to exhibit longer bonds (2.02 – 2.22 Å) than [Ag$^{III}$F$_4$]$^-$ squares (1.91–1.94 Å). No spin-density is found at the [Ag$^{III}$F$_4$]$^-$ units, in accordance with the low-spin nature of square-coordinated $d^8$ complexes.

Upon compression to 3 GPa a new $P\bar{1}$ ($Z = 4$) structure, dubbed $P\bar{1}(\beta_1)$, becomes the most stable polymorph. This structure also contains Ag$^{3+}$ and Ag$^{2+}$ cations, but in contrast to $P\bar{1}(\alpha)$ the latter are found in two different coordination environments. Three out of the four Ag$^{2+}$ sites feature compressed octahedral geometry with two short Ag-F bonds (2.03 – 2.08 Å at 5 GPa) and four longer contacts (2.30 – 2.52 Å). The fourth site adopts the geometry of an elongated octahedron with four short bonds (2.02 – 2.08 Å), forming a square around $Ag^{2+}$, and two longer ones (2.60 Å) in trans positions. Sites containing $Ag^{3+}$ retain the square-planar geometry of $P\bar{1}(\alpha)$ with Ag-F bonds between 1.92 and 1.96 Å, as expected. We note that the compressed octahedral coordination of ions with a $d^9$ electron count pushes the spin density into an orbital resembling the $d(z^2)$ orbital (where z is the axis of compression) of $A_{1g}$ symmetry. On the other hand, elongation results in the unpaired electron occupying a $d(x^2-y^2)$-like orbital of $B_{1g}$ symmetry.[71]

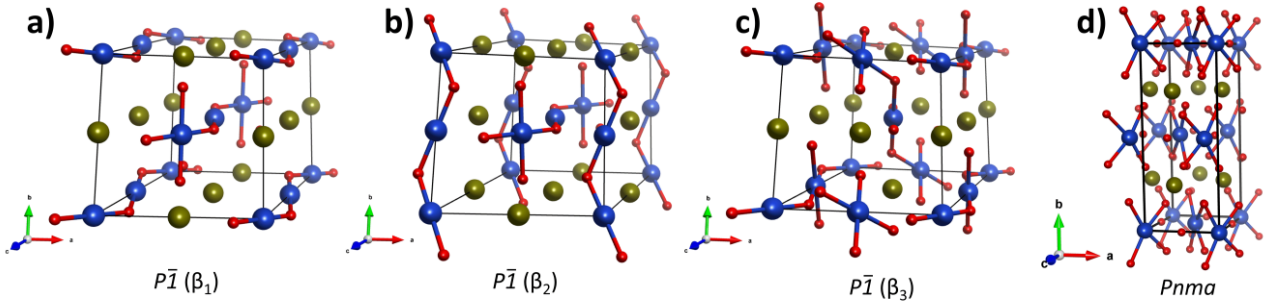

**Figure 5** The structures of $Ag_2F_5$ polymorphs: $P\bar{1}(\beta_1)$ (a), $P\bar{1}(\beta_2)$ (b), $P\bar{1}(\beta_3)$ (c), and *Pnma* (d). Blue/gold/red balls depict $Ag^{II}/Ag^{III}/F$ atoms.

The connectivity of $Ag^{2+}$ sites in $P\bar{1}(\beta_1)$ result in the formation of two types of chains containing $Ag^{2+}$ cations (Figure 5a), both running along the **c** lattice vector. One is built from compressed octahedra joined by short Ag-F bonds, the other features alternation of compressed/elongated sites. The former structural motif (which we refer to as an $A_{1g}$ chain) is found at ambient conditions in a few compounds of $Ag^{2+}$ and $Cu^{2+}$.[40] Due to the good overlap between the $d(z^2)$-type orbitals of neighbouring $Ag^{2+}$ sites such chains host strong AFM superexchange interactions.[21] The other type of chain, which we call $A_{1g}/B_{1g}$ chain, has not been not reported to date.

The two structures that follow $P\bar{1}(\beta_1)$ in the pressure-induced phase transition sequence are very similar to it (Figure 5b,c). These are $P\bar{1}(\beta_2)$, stable from 7 to 36 GPa, and $P\bar{1}(\beta_3)$, which is the ground state structure from 36 to 65 GPa. All three have Ag atoms at the same fractional positions and possess very similar lattice vectors. They differ in the positions of the fluorine atoms, and hence in the arrangement of $Ag^{2+}/Ag^{3+}$ sites, which results in changes in the $Ag^{2+}$ framework connectivity. In $P\bar{1}(\beta_2)$ we have $A_{1g}/B_{1g}$ chains running along the **c** lattice vector, and $A_{1g}$ chains along the **b** vector (Figure 5b). In contrast, $P\bar{1}(\beta_3)$ exhibits only $A_{1g}/B_{1g}$ chains (half of the $Ag^{2+}$ sites possess an elongated octahedral coordination) running along the [001] and [011] crystallographic directions. As will be shown below, all three β-type monoclinic phases exhibit strong AFM coupling within both types of $Ag^{2+}$ chains, and hence are predicted to exhibit quasi-1D magnetic properties.

Above 65 GPa, $Ag_2F_5$ is predicted to adopt a structure of *Pnma* symmetry ($Z = 4$) containing $Ag^{2+}$ in a highly distorted square environment (Figure 5d). This structure bears some resemblance to the $KBrF_4$ polytype (containing $BrF_4^-$ squares separated by $K^+$ cations),[72] but with additional F atoms located between the $[AgF_4]^{2-}$ plaquettes. $Ag^{II}Ag^{III}F_5$ remains in the *Pnma* structure up to the limit of its thermodynamic stability at 83 GPa – above that pressure this compound should decompose to a mixture of $Ag^{II}F_2$ and $Ag^{III}F_3$.

## $Ag_3F_8$

The second $Ag^{2+}/Ag^{3+}$ mixed-valent fluoride known experimentally, $Ag_3F_8$ (= $Ag^{II}[Ag^{III}F_4]_2$), crystallizes at ambient pressure in a structure of $P2_1/n$ symmetry ($Z = 2$) consisting of isolated (*i.e.* non-bridging) $[AgF_4]^{2-}$ units extending in the **ab** plane (Figure 6a).[51] We predict that at 1.5 GPa this structure should transform to a related phase of $P2_1/c$ symmetry ($Z = 2$), shown in Figure 6b. This phase differs from $P2_1/n$ in the stacking of $Ag^{2+}$ sites along the *c* axis (in $P2_1/c$ the atoms at $z = ½$ are translated by ½ in the direction of the *a* vector). Moreover, in $P2_1/c$ $Ag^{2+}$ cations are found in a compressed octahedral coordination leading to the formation of $AgF_2$ dumbbells. This indicates that $Ag_3F_8$ should exhibit a pressure-induced reversal of the direction of the Jahn-Teller effect, testifying to the large plasticity of the first coordination sphere of $d^9$ cations ($Cu^{2+}$, $Ag^{2+}$).[73,74] $Ag^{II}[Ag^{III}F_4]_2$ becomes thermodynamically unstable above 18 GPa, when it is predicted to disproportionate into $Ag_2F_5$ and $AgF_3$.

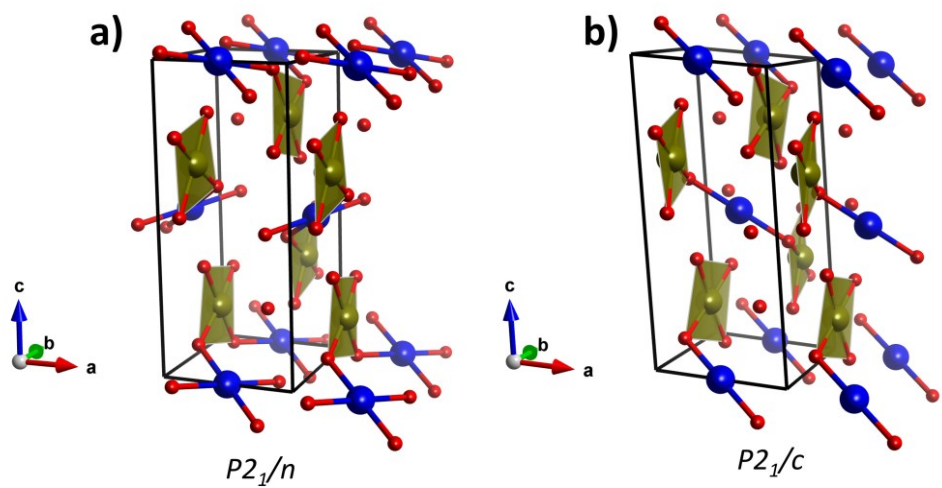

**Figure 6** The crystal structure of the most stable phases of $Ag_3F_8$: $P2_1/n$ (a), and $P2_1/c$ (b). Blue/gold/red balls depict $Ag^{II}/Ag^{III}/F$ atoms.

Comparing the $Ag^{II}$ network in all of the five compounds containing silver in this oxidation state ($Ag_3F_4$, $Ag_2F_3$, $AgF_2$, $Ag_2F_5$, $Ag_3F_8$) we find that only the low-pressure *Pbca* phase of $AgF_2$ exhibits a 2D network, and consequently quasi-2D magnetic properties.[20] The other fluorides exhibit $Ag^{II}$ centres either joined in chains ($Ag_2F_3$, $P\bar{1}(\beta)$ phases of $Ag_2F_5$) – for which quasi-1D magnetic properties as expected; or isolated sites ($Ag_3F_4$, $Ag_3F_5$), which should lead to negligible magnetic coupling. The magnetic spin-spin interactions in the $P\bar{1}(\beta)$ phases of $Ag_2F_5$ will be described in details further in the text.

It seems that pressure acts to reduce the dimensionality of the $Ag^{II}$ network with a 1D to 0D (*Pnma* phase) transition in $Ag_2F_5$, and a 2D to 1D (*Pca2_1* phase) to 0D (*Pbcn* phase) transition in $AgF_2$, as reported earlier.[18–20]

**Silver tri- and tetrafluoride ($AgF_3$, $AgF_4$)**

**$AgF_3$**

At atmospheric pressure, $AgF_3$ adopts an $AuF_3$-type structure featuring helical chains built from $[AgF_4]^-$ squares sharing two *cis* fluorine atoms. Recent computational studies predicted that upon compression this structure should transform into a polymorph with $P\bar{1}$ symmetry ($Z = 2$) featuring *trans* bridges (Figure 7a).[3] This polymorph, labelled as $P\bar{1}(\alpha)$, should subsequently transform into a $P\bar{1}(\beta)$ structure exhibiting double-bridged $Ag_2F_6$ units. Our structure searches yielded two novel structures: $P2_1$ ($Z = 4$) built from trans-connected chains (Figure 7b), and $P\bar{1}(\gamma)$ ($Z = 6$), which is a layered structure (Figure 7c,d). Consequently, we find the following phase transition sequence up to 100 GPa: $P6_122 \xrightarrow{3\,GPa} P2_1 \xrightarrow{41\,GPa} P\bar{1}(\alpha) \xrightarrow{76\,GPa} P\bar{1}(\gamma)$. In all ground-state structures, $Ag^{3+}$ retains the square planar coordination. In contrast to previous results, we do not find any pressure range where $P\bar{1}(\beta)$ is stable. Instead, the formation of the layered $P\bar{1}(\gamma)$ structure is favored. We note that despite large differences in the bonding framework between $P\bar{1}(\alpha)$ and $P\bar{1}(\gamma)$ both polymorphs differ by less than 25 meV/$AgF_3$ in the 40–100 GPa pressure range. The chemical formula of the $P\bar{1}(\gamma)$ phase might be approximated as $[Ag^{III}_5F_{14}]^+[Ag^{III}F_4]^-$; a similar autodissociation has been observed at elevated pressure for a number of compounds including $XeF_2$ and $NH_3$.[75–77]

Our findings indicate that for $AgF_3$ compression leads to a transition from chain structures ($P6_122, P2_1, P\bar{1}(\alpha)$) to a layered polymorph, in contrast to what is found for $AuF_3$, which is predicted to adopt the dimeric $P\bar{1}(\beta)$ structure at the highest pressures studied.[3,4,28] These difference might be a result of the strong relativistic effects found for Au,[78–80] although this hypothesis should be verified by future calculations that are beyond the scope of this work.

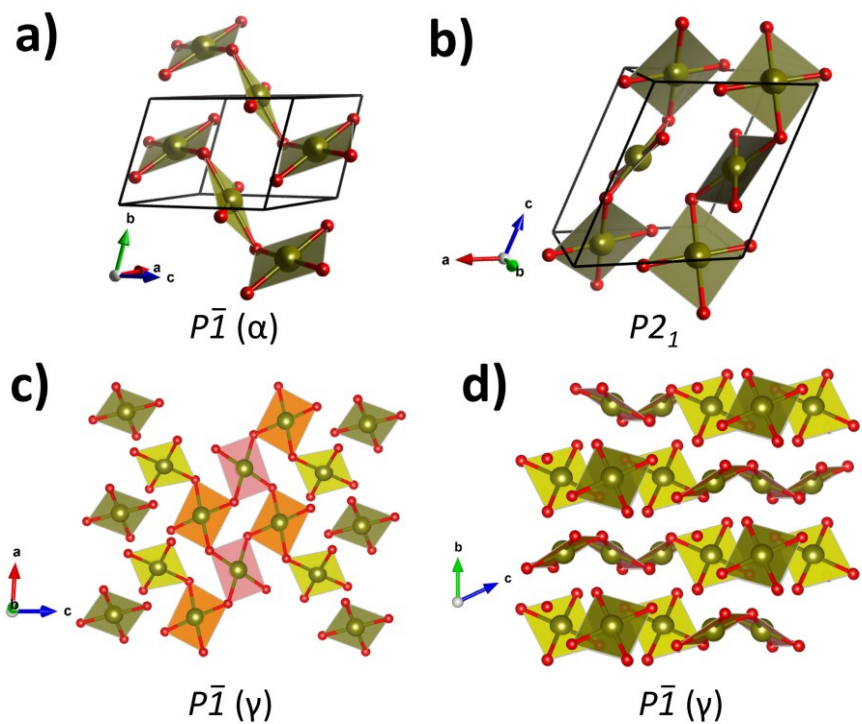

**Figure 7** The crystal structure of the most stable phases of AgF$_3$: $P\bar{1}(\alpha)$ (a), P2$_1$ (b), and $P\bar{1}(\gamma)$ (c,d). Gold/red balls depict Ag$^{III}$/F atoms. In case of $P\bar{1}(\gamma)$ different colours are given for AgF$_4$ squares featuring different connectivity (gold – isolated, yellow/orange/pink – sharing one/two/three F atoms). Note, the chemical formula of the $P\bar{1}(\gamma)$ phase might be approximated as (Ag$_5$F$_{14}$)$^+$(AgF$_4$)$^-$.

**AgF$_4$**

Results of a recent computational study (utilizing the DFT GGA method) indicated that silver tetrafluoride should become thermodynamically stable at 37.5 GPa forming a tetragonal (*I4/m*, *Z* = 2) structure.³ Our meta-GGA calculations place the onset of thermodynamic stability of AgF$_4$ at 56 GPa. We find that at this pressure AgF$_4$ should form a *C2/m* (*Z = 2*) molecular structure, which can be obtained from the *I4/m* polymorph through a distortion that leads to the alternation of the length of the

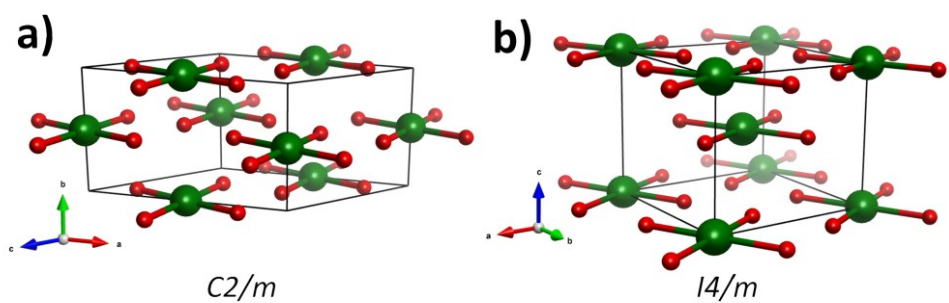

**Figure 8** The crystal structure of the most stable phases of AgF$_4$: *C2/m* (a), and *I4/m* (b). Green balls depict Ag$^{IV}$ atoms.

*a* and *b* lattice vectors, and departure of the angle between the two vectors from 90°. Upon compression, *C2/m* smoothly transforms towards *I4/m* and is identical to that structure above 69 GPa.

Both AgF$_4$ polymorphs are molecular crystals containing square planar AgF$_4$ units with short Ag-F bonds (1.88 Å at 60 GPa). For comparison, the length of the bridging/terminal bonds in the $P\bar{1}(\alpha)$ polymorph of AgF$_3$ at 60 GPa is 1.95/1.88 Å. Our calculations indicate that AgF$_4$ should exhibit paramagnetic properties with the total unpaired electron density equal to 0.95μ$_B$ per AgF$_4$ – in accordance with an assignment of a +4 oxidation state for Ag (resulting in one unpaired electron per Ag site). We note, however, that the spin density is equally distributed within the whole AgF$_4$ unit with the magnetization of Ag/F being equal to 0.23μ$_B$/0.18μ$_B$. This testifies to the large covalence of the Ag$^{IV}$–F bonds, which certainly exceeds those found for Ag$^{III}$ and Ag$^{II}$ analogues.[81] Moreover, both polymorphs of AgF$_4$ are metallic, as described below.

**Superexchange interactions in Ag$_2$F$_5$**

For the three types of $P\bar{1}(\beta)$ polymorphs of Ag$_2$F$_5$ we have evaluated the AFM superexchange coupling constants for the A$_{1g}$ and A$_{1g}$/B$_{1g}$ chains. We assumed the Heisenberg Hamiltonian in the form $\boldsymbol{H}_{ij}$ = –$J_{ij}·\boldsymbol{s}_i·\boldsymbol{s}_j$, which results in negative $J_{ij}$ values for the antiferromagnetic (AFM) coupling – the more negative the stronger the coupling. The resulting coupling constants, extracted with the use of the broken-symmetry method[38,39] (as applied in refs. [20,40]), are shown in Figure 9 as a function of the angle of the Ag-F-Ag bridge. We note that the interchain coupling constants are computed to be smaller than the intrachain ones by at least an order of magnitude, which testifies to the quasi-1D magnetic character of the $P\bar{1}(\beta)$ polymorphs of $Ag_2F_5$.

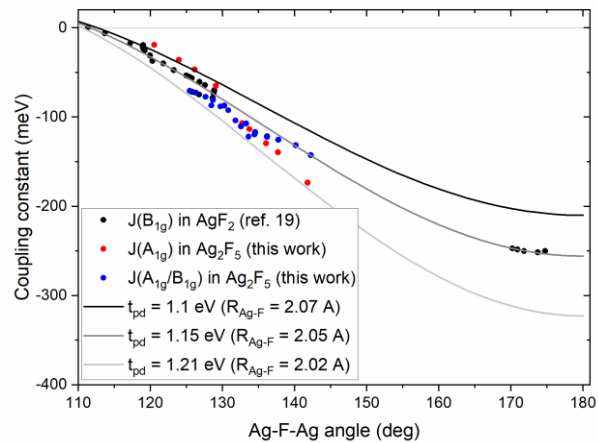

**Figure 9** The dependence of the values of the magnetic coupling constants for $P\bar{1}(\beta)$ polymorphs of Ag$_2$F$_5$ on the angle of the Ag–F–Ag bridge (red/blue circles for A$_{1g}$ and A$_{1g}$/B$_{1g}$ chains – see text). The black points are coupling constant values calculated for the high-pressure polymorphs of AgF$_2$ (featuring sheets, chains, and dimers with Ag$^{II}$ centres in the B$_{1g}$ electronic state) taken from ref. [20] Full lines show the result of an analytical computation (also taken from ref. [20]) of *J* for a single Ag–F–Ag bridge (assuming a B$_{1g}$ electronic state) with different values of $t_{pd}$ hopping integrals, which correspond to different Ag-F bond lengths.

For both types of bridges, the strength of the AFM interaction decreases with the decrease of the linearity of the Ag-F-Ag bridge, as expected within the framework of the Goodenough-Kanamori rules.[82–84] The dependence on the angle is more steep for the $A_{1g}$ chains, which may be attributed to the better overlap between the ligand orbitals and the spin-carrying orbital in case of $A_{1g}$-type coordination compared to $B_{1g}$-type,[21] as rationalized in the orbital description of the superexchange interaction introduced by Hoffman and co-workers.[85]

As can be seen in Figure 9 the angle dependence of the coupling constants calculated for $Ag_2F_5$ is similar to that predicted for the high-pressure polymorphs of $AgF_2$.[20] We note that all of the structures of $AgF_2$ exhibit $Ag^{II}$ centres in the $B_{1g}$ electronic state, hence the superexchange pathways are of $B_{1g}$ type. As compression leads to larger bending of the Ag-F-Ag bridges in all three $Ag_2F_5$ structures, pressure leads to weaken the AFM interactions. The strongest AFM coupling ($J = -173$ meV and $-143$ meV for the $A_{1g}$ and $A_{1g}/B_{1g}$ chains) is predicted to occur for $P\bar{1}(\beta_1)$ at 1 atm. This can be compared with the value of $-71$ meV derived with the same method and functional for the 2D AFM coupled-lattice of $AgF_2$.[20]

**Electronic structure**

Table 1 gives the electronic band gap ($E_g$) for $Ag_mF_n$ compounds calculated with the SCAN and HSE06 functional at selected pressures within their thermodynamic stability range. The HSE06 method yields band gaps closer to experimental values, and therefore we will use the values obtained with this method while discussing the electronic properties of $Ag_mF_n$.

At pressure up to 10 GPa, compounds with $Ag^I/Ag^{II}$ mixed-valence ($Ag_3F_4$, $Ag_2F_3$) are predicted to exhibit a finite band gap of around 1.6 eV – somewhat smaller than the value calculated for $AgF_2$ using the same methodology (2.3 eV).

For $Ag^{II}/Ag^{III}$ fluorides, the values of the band gap are smaller than those calculated for $Ag^{II}F_2$ and $Ag^{III}F_3$, but again finite. The band gap of these compounds, as well as that of $AgF_3$, is predicted to diminish upon compression (Figure 10) – in contrast to what was predicted for $AgF_2$.[20] Interestingly, silver tetrafluoride is predicted to be a ferromagnetic half-metal with conductivity only in the spin minority (spin down) channel (Figure 10). The density of states at the Fermi level has $F(p)$ character, hinting that the metallic behaviour might be result of the compression-induced orbital overlap between neighbouring $AgF_4$ molecules.

**Table 1** The electronic band gap ($E_g$, given in eV) for selected structures in the Ag/F system as calculated with the HSE06 functional for the SCAN-optimized structures. The band gap values derived with SCAN are given in parentheses. Pressure (P) is given in GPa. The abbreviation 'met.' denotes metallic behavior is indicated by (no band gap).

| System | Structure | P | $E_g$ |
|---|---|---|---|
| $Ag_3F_4$ (= $Ag^I_2Ag^{II}F_4$) | $I\bar{4}2d$ | 10 | 1.63 (0.30) |
| $Ag_2F_3$ (= $Ag^IAg^{II}F_3$) | $P2_1/m$ | 0 | 1.54 (0.24) |
| $Ag^{II}F_2$ | $Pbca$ | 0 | 2.27 (0.60) |
|  | $Pca2_1$ | 10 | 2.29 (0.64) |
| $Ag_2F_5$ (= $Ag^{II}F[Ag^{III}F_4]$) | $P\bar{1}(\beta_2)$ | 10 | 1.64 (0.26) |
|  | $P\bar{1}(\beta_3)$ | 60 | 1.21 (0.12) |
| $Ag_3F_8$ (= $Ag^{II}[Ag^{III}F_4]_2$) | $P2_1/c$ | 10 | 1.98 (0.52) |
| $Ag^{III}F_3$ | $P2_1$ | 10 | 2.22 (0.82) |
|  | $P\bar{1}(\alpha)$ | 60 | 1.75 (0.32) |
|  | $P\bar{1}(\gamma)$ | 80 | 0.91 (0.05) |
| $Ag^{IV}F_4$ | $C2/m$ | 60 | met. (met.) |
|  | $I4/m$ | 80 | met. (met.) |

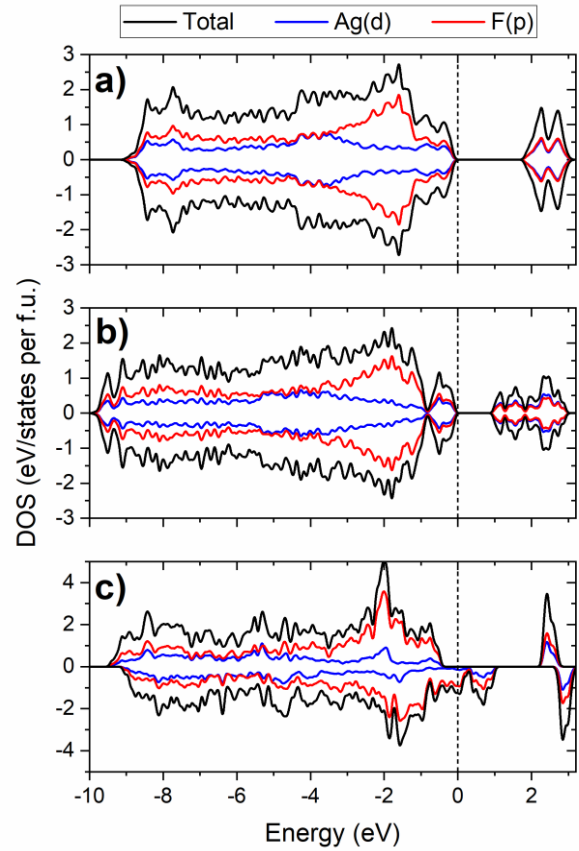

**Figure 10** Total (black line) and partial (blue/red line for Ag(*d*)/F(*p*)) electronic DOS calculated with HSE06 for: $AgF_3$ in the $P\bar{1}(\alpha)$ structure at 60 GPa (a); $AgF_3$ in the $P\bar{1}(\gamma)$ structure at 80 GPa (b); $AgF_4$ in the $I4/m$ structure at 80 GPa (c).

## Conclusions

In conclusion, our computational study shows that two novel fluorides of silver, $Ag_3F_4$ (= $Ag^I_2Ag^{II}F_4$), and $Ag_2F_3$ (= $Ag^IAg^{II}F_3$) are thermodynamically stable at ambient conditions. The former should adopt a distorted spinel structure that corresponds to the most stable high-pressure polymorph for a number of $ABX_3$ systems. Compression leads to destabilization of both $Ag^+/Ag^{2+}$ and $Ag^{2+}/Ag^{3+}$ mixed-valent fluorides, although $Ag_2F_5$ (= $Ag^{II}F[Ag^{III}F_4]$) remains on the convex hull of the $Ag/F_2$ system up to 83 GPa. In this system we predict a sequence of phase transition between 3 and 65 GPa between closely related structures differing mainly in the position of the fluorine atoms. Changes in the geometry of the fluorine sub-lattice leads to switching of the positions of $Ag^{2+}/Ag^{3+}$ sites, but the 1D character of the AFM superexchange interactions remains intact. For the $Ag_3F_8$ system (= $Ag^{II}[Ag^{III}F_4]$) we predict a pressure-induced switching of the Jahn-Teller distortion from an elongated to a compressed octahedron. Finally, we confirm the paramagnetic and metallic nature of $AgF_4$ while showing that the spin density in this compound is almost evenly distributed among $Ag^{2+}$ cations and $F^-$ anions.

## Acknowledgements

The authors acknowledge Polish National Science Centre (Maestro grant 2017/26/A/ST5/00570) and the Interdisciplinary Centre for Mathematical and Computational Modelling (grants GB80-11 and GA67-13). E. Z. acknowledges the NSF (DMR-1827815) for financial support, and the Center for Computational Research (CCR) at SUNY Buffalo for computational support.[86] M. D. acknowledges The European Regional Development Fund, Research and Innovation Operational Programme (project No. ITMS2014+: 313011W085), Scientific Grant Agency of the Slovak Republic (grant No. VG 1/0223/19) and the Slovak Research and Development Agency (grant No. APVV-18-0168) and Aurel supercomputing infrastructure in CC of Slovak Academy of Sciences acquired in projects ITMS 26230120002 and 26210120002 funded by ERDF.

# Electronic Supplementary Information

## Fluorides of silver at large compression

Dominik Kurzydłowski,[a*] Mariana Derzsi,[b,c] Eva Zurek,[d] Wojciech Grochala[c]

Cif files of selected silver fluorides:

**Ag$_2$F *I4/mmm* @ 60 GPa**
```
data_findsym-output
_audit_creation_method FINDSYM

_cell_length_a    2.6112400000
_cell_length_b    2.6112400000
_cell_length_c    9.6611896324
_cell_angle_alpha 90.0000000000
_cell_angle_beta  90.0000000000
_cell_angle_gamma 90.0000000000
_cell_volume      65.8755396978

_symmetry_space_group_name_H-M "I 4/m 2/m 2/m"
_symmetry_Int_Tables_number 139
_space_group.reference_setting '139:-I 4 2'
_space_group.transform_Pp_abc a,b,c;0,0,0

loop_
_space_group_symop_id
_space_group_symop_operation_xyz
1 x,y,z
2 x,-y,-z
3 -x,y,-z
4 -x,-y,z
5 -y,-x,-z
6 -y,x,z
7 y,-x,z
8 y,x,-z
9 -x,-y,-z
10 -x,y,z
11 x,-y,z
12 x,y,-z
13 y,x,z
14 y,-x,-z
15 -y,x,-z
16 -y,-x,z
17 x+1/2,y+1/2,z+1/2
18 x+1/2,-y+1/2,-z+1/2
19 -x+1/2,y+1/2,-z+1/2
20 -x+1/2,-y+1/2,z+1/2
21 -y+1/2,-x+1/2,-z+1/2
22 -y+1/2,x+1/2,z+1/2
23 y+1/2,-x+1/2,z+1/2
24 y+1/2,x+1/2,-z+1/2
25 -x+1/2,-y+1/2,-z+1/2
26 -x+1/2,y+1/2,z+1/2
```

```
27 x+1/2,-y+1/2,z+1/2
28 x+1/2,y+1/2,-z+1/2
29 y+1/2,x+1/2,z+1/2
30 y+1/2,-x+1/2,-z+1/2
31 -y+1/2,x+1/2,-z+1/2
32 -y+1/2,-x+1/2,z+1/2

loop_
_atom_site_label
_atom_site_type_symbol
_atom_site_symmetry_multiplicity
_atom_site_Wyckoff_label
_atom_site_fract_x
_atom_site_fract_y
_atom_site_fract_z
_atom_site_occupancy
_atom_site_fract_symmform
Ag1 Ag   4 e 0.00000 0.00000 0.15193 1.00000 0,0,Dz
F1  F    2 b 0.00000 0.00000 0.50000 1.00000 0,0,0
```

**Ag2F *P4/mmm* @ 60 GPa**

```
data_findsym-output
_audit_creation_method FINDSYM

_cell_length_a    2.6505700000
_cell_length_b    2.6505700000
_cell_length_c    9.3518700000
_cell_angle_alpha 90.0000000000
_cell_angle_beta  90.0000000000
_cell_angle_gamma 90.0000000000
_cell_volume      65.7017621127

_symmetry_space_group_name_H-M "P 4/m 2/m 2/m"
_symmetry_Int_Tables_number 123
_space_group.reference_setting '123:-P 4 2'
_space_group.transform_Pp_abc a,b,c;0,0,0

loop_
_space_group_symop_id
_space_group_symop_operation_xyz
1 x,y,z
2 x,-y,-z
3 -x,y,-z
4 -x,-y,z
5 -y,-x,-z
6 -y,x,z
7 y,-x,z
8 y,x,-z
9 -x,-y,-z
10 -x,y,z
11 x,-y,z
12 x,y,-z
13 y,x,z
14 y,-x,-z
15 -y,x,-z
16 -y,-x,z

loop_
_atom_site_label
_atom_site_type_symbol
```

_atom_site_symmetry_multiplicity
_atom_site_Wyckoff_label
_atom_site_fract_x
_atom_site_fract_y
_atom_site_fract_z
_atom_site_occupancy
_atom_site_fract_symmform
Ag1 Ag   2 h 0.50000 0.50000 0.69920 1.00000 0,0,Dz
Ag2 Ag   1 c 0.50000 0.50000 0.00000 1.00000 0,0,0
Ag3 Ag   1 b 0.00000 0.00000 0.50000 1.00000 0,0,0
F1  F    2 g 0.00000 0.00000 0.85020 1.00000 0,0,Dz

**Ag$_3$F$_4$ *I-42d* @ 10 GPa**
data_findsym-output
_audit_creation_method FINDSYM

_cell_length_a    6.7087400000
_cell_length_b    6.7087400000
_cell_length_c    7.4785200000
_cell_angle_alpha 90.0000000000
_cell_angle_beta  90.0000000000
_cell_angle_gamma 90.0000000000
_cell_volume      336.5871884145

_symmetry_space_group_name_H-M "I -4 2 d"
_symmetry_Int_Tables_number 122
_space_group.reference_setting '122:I -4 2bw'
_space_group.transform_Pp_abc a,b,c;0,0,0

loop_
_space_group_symop_id
_space_group_symop_operation_xyz
1 x,y,z
2 x,-y+1/2,-z+1/4
3 -x,y+1/2,-z+1/4
4 -x,-y,z
5 y,x+1/2,z+1/4
6 y,-x,-z
7 -y,x,-z
8 -y,-x+1/2,z+1/4
9 x+1/2,y+1/2,z+1/2
10 x+1/2,-y,-z+3/4
11 -x+1/2,y,-z+3/4
12 -x+1/2,-y+1/2,z+1/2
13 y+1/2,x,z+3/4
14 y+1/2,-x+1/2,-z+1/2
15 -y+1/2,x+1/2,-z+1/2
16 -y+1/2,-x,z+3/4

loop_
_atom_site_label
_atom_site_type_symbol
_atom_site_symmetry_multiplicity
_atom_site_Wyckoff_label
_atom_site_fract_x
_atom_site_fract_y
_atom_site_fract_z
_atom_site_occupancy
_atom_site_fract_symmform
Ag1 Ag   8 d 0.40805 0.25000 0.12500 1.00000 Dx,0,0

Ag2 Ag   4 a 0.00000 0.00000 0.00000 1.00000 0,0,0
F1  F   16 e 0.30599 0.57027 0.20870 1.00000 Dx,Dy,Dz

**Ag$_2$F$_3$ *P2$_1$/m* @ 0 GPa**
data_findsym-output
_audit_creation_method FINDSYM

_cell_length_a    3.3460700000
_cell_length_b    7.6103000000
_cell_length_c    5.4786200000
_cell_angle_alpha 90.0000000000
_cell_angle_beta  99.5383300000
_cell_angle_gamma 90.0000000000
_cell_volume      137.5821056504

_symmetry_space_group_name_H-M "P 1 21/m 1"
_symmetry_Int_Tables_number 11
_space_group.reference_setting '011:-P 2yb'
_space_group.transform_Pp_abc a,b,c;0,0,0

loop_
_space_group_symop_id
_space_group_symop_operation_xyz
1 x,y,z
2 -x,y+1/2,-z
3 -x,-y,-z
4 x,-y+1/2,z

loop_
_atom_site_label
_atom_site_type_symbol
_atom_site_symmetry_multiplicity
_atom_site_Wyckoff_label
_atom_site_fract_x
_atom_site_fract_y
_atom_site_fract_z
_atom_site_occupancy
_atom_site_fract_symmform
Ag1 Ag   2 b 0.50000 0.00000  0.00000 1.00000 0,0,0
Ag2 Ag   2 e 0.21855 0.25000  0.51087 1.00000 Dx,0,Dz
F1  F    2 e 0.63468 0.25000  0.16428 1.00000 Dx,0,Dz
F2  F    4 f 0.18856 -0.05864 0.29123 1.00000 Dx,Dy,Dz

**Ag$_2$F$_5$ *P-1(β$_1$)* @ 10 GPa**
data_findsym-output
_audit_creation_method FINDSYM

_cell_length_a    7.0919930000
_cell_length_b    7.2427570000
_cell_length_c    7.5178760000
_cell_angle_alpha 117.9999770000
_cell_angle_beta  112.6565090000
_cell_angle_gamma 90.2877120000
_cell_volume      306.3522431633

_symmetry_space_group_name_H-M "P -1"
_symmetry_Int_Tables_number 2
_space_group.reference_setting '002:-P 1'
_space_group.transform_Pp_abc a,b,c;0,0,0

```
loop_
_space_group_symop_id
_space_group_symop_operation_xyz
1 x,y,z
2 -x,-y,-z

loop_
_atom_site_label
_atom_site_type_symbol
_atom_site_symmetry_multiplicity
_atom_site_Wyckoff_label
_atom_site_fract_x
_atom_site_fract_y
_atom_site_fract_z
_atom_site_occupancy
_atom_site_fract_symmform
Ag1 Ag   1 a 0.00000 0.00000  0.00000 1.00000 0,0,0
Ag2 Ag   1 b 0.00000 0.00000  0.50000 1.00000 0,0,0
Ag3 Ag   1 e 0.50000 0.50000  0.00000 1.00000 0,0,0
Ag4 Ag   1 h 0.50000 0.50000  0.50000 1.00000 0,0,0
Ag5 Ag   1 c 0.00000 0.50000  0.00000 1.00000 0,0,0
Ag6 Ag   1 g 0.00000 0.50000  0.50000 1.00000 0,0,0
Ag7 Ag   1 d 0.50000 0.00000  0.00000 1.00000 0,0,0
Ag8 Ag   1 f 0.50000 0.00000  0.50000 1.00000 0,0,0
F1  F    2 i 0.29992 0.81570  0.49686 1.00000 Dx,Dy,Dz
F2  F    2 i 0.09036 -0.05601 0.75700 1.00000 Dx,Dy,Dz
F3  F    2 i 0.61783 0.49054  0.79307 1.00000 Dx,Dy,Dz
F4  F    2 i 0.12852 0.38862  0.30117 1.00000 Dx,Dy,Dz
F5  F    2 i 0.69891 0.82034  0.49502 1.00000 Dx,Dy,Dz
F6  F    2 i 0.35401 0.13673  0.17827 1.00000 Dx,Dy,Dz
F7  F    2 i 0.56456 0.82730  0.14040 1.00000 Dx,Dy,Dz
F8  F    2 i 0.16591 0.77037  0.09467 1.00000 Dx,Dy,Dz
F9  F    2 i 0.71941 0.37404  0.26941 1.00000 Dx,Dy,Dz
F10 F    2 i 0.10392 0.34228  0.77801 1.00000 Dx,Dy,Dz
```

**$Ag_2F_5$ *Pnma* @ 50 GPa**
data_findsym-output
_audit_creation_method FINDSYM

_cell_length_a    4.4881320000
_cell_length_b    11.0119560000
_cell_length_c    4.9715920000
_cell_angle_alpha 90.0000000000
_cell_angle_beta  90.0000000000
_cell_angle_gamma 90.0000000000
_cell_volume      245.7115487622

_symmetry_space_group_name_H-M "P 21/n 21/m 21/a"
_symmetry_Int_Tables_number 62
_space_group.reference_setting '062:-P 2ac 2n'
_space_group.transform_Pp_abc a,b,c;0,0,0

loop_
_space_group_symop_id
_space_group_symop_operation_xyz
1 x,y,z
2 x+1/2,-y+1/2,-z+1/2
3 -x,y+1/2,-z
4 -x+1/2,-y,z+1/2
5 -x,-y,-z

6 -x+1/2,y+1/2,z+1/2
7 x,-y+1/2,z
8 x+1/2,y,-z+1/2

loop_
_atom_site_label
_atom_site_type_symbol
_atom_site_symmetry_multiplicity
_atom_site_Wyckoff_label
_atom_site_fract_x
_atom_site_fract_y
_atom_site_fract_z
_atom_site_occupancy
_atom_site_fract_symmform
Ag1 Ag   4 c 0.40996 0.25000 -0.00797 1.00000 Dx,0,Dz
Ag2 Ag   4 a 0.00000 0.00000  0.00000 1.00000 0,0,0
F1  F    8 d 0.29134 0.88237  0.71253 1.00000 Dx,Dy,Dz
F2  F    8 d 0.45140 -0.08411 0.16543 1.00000 Dx,Dy,Dz
F3  F    4 c 0.40241 0.25000  0.53920 1.00000 Dx,0,Dz

**$Ag_3F_8$ *$P2_1/c$* @ 10 GPa**
data_findsym-output
_audit_creation_method FINDSYM

_cell_length_a    4.6415700000
_cell_length_b    5.2518900000
_cell_length_c   10.6572300000
_cell_angle_alpha 90.0000000000
_cell_angle_beta  96.9586400000
_cell_angle_gamma 90.0000000000
_cell_volume     257.8777977703

_symmetry_space_group_name_H-M "P 1 21/c 1"
_symmetry_Int_Tables_number 14
_space_group.reference_setting '014:-P 2ybc'
_space_group.transform_Pp_abc a,b,c;0,0,0

loop_
_space_group_symop_id
_space_group_symop_operation_xyz
1 x,y,z
2 -x,y+1/2,-z+1/2
3 -x,-y,-z
4 x,-y+1/2,z+1/2

loop_
_atom_site_label
_atom_site_type_symbol
_atom_site_symmetry_multiplicity
_atom_site_Wyckoff_label
_atom_site_fract_x
_atom_site_fract_y
_atom_site_fract_z
_atom_site_occupancy
_atom_site_fract_symmform
Ag1 Ag   2 d 0.50000 0.00000 0.50000 1.00000 0,0,0
Ag2 Ag   4 e 0.22380 0.45844 0.68466 1.00000 Dx,Dy,Dz
F1  F    4 e 0.25017 0.62869 0.52775 1.00000 Dx,Dy,Dz
F2  F    4 e 0.24136 0.28202 0.84517 1.00000 Dx,Dy,Dz
F3  F    4 e 0.18018 0.37039 0.09845 1.00000 Dx,Dy,Dz

F4  F    4 e 0.28241 0.73477 0.28303 1.00000 Dx,Dy,Dz

**AgF$_3$ *P2$_1$* @ 20 GPa**
data_findsym-output
_audit_creation_method FINDSYM

_cell_length_a    5.7068510000
_cell_length_b    4.6209180000
_cell_length_c    6.5574900000
_cell_angle_alpha 90.0000000000
_cell_angle_beta  112.0484160000
_cell_angle_gamma 90.0000000000
_cell_volume      160.2801869345

_symmetry_space_group_name_H-M "P 1 21 1"
_symmetry_Int_Tables_number 4
_space_group.reference_setting '004:P 2yb'
_space_group.transform_Pp_abc a,b,c;0,0,0

loop_
_space_group_symop_id
_space_group_symop_operation_xyz
1 x,y,z
2 -x,y+1/2,-z

loop_
_atom_site_label
_atom_site_type_symbol
_atom_site_symmetry_multiplicity
_atom_site_Wyckoff_label
_atom_site_fract_x
_atom_site_fract_y
_atom_site_fract_z
_atom_site_occupancy
_atom_site_fract_symmform
Ag1 Ag   2 a 0.25004  0.87632  0.00000 1.00000 Dx,Dy,Dz
Ag2 Ag   2 a 0.25000  0.87647  0.50002 1.00000 Dx,Dy,Dz
F1  F    2 a 0.38427  0.69086  0.29695 1.00000 Dx,Dy,Dz
F2  F    2 a 0.45230  0.33521  0.05155 1.00000 Dx,Dy,Dz
F3  F    2 a 0.76431  0.04043  0.34360 1.00000 Dx,Dy,Dz
F4  F    2 a 0.73582  0.71253  0.65634 1.00000 Dx,Dy,Dz
F5  F    2 a -0.04762 -0.08243 0.05154 1.00000 Dx,Dy,Dz
F6  F    2 a 0.11586  0.06210  0.70317 1.00000 Dx,Dy,Dz

**AgF$_3$ *P-1(γ)* @ 60 GPa**
data_findsym-output
_audit_creation_method FINDSYM

_cell_length_a    5.2308600000
_cell_length_b    6.1492600000
_cell_length_c    6.8420400000
_cell_angle_alpha 64.5152200000
_cell_angle_beta  88.1154800000
_cell_angle_gamma 87.0047100000
_cell_volume      198.3819615231

_symmetry_space_group_name_H-M "P -1"
_symmetry_Int_Tables_number 2
_space_group.reference_setting '002:-P 1'
_space_group.transform_Pp_abc a,b,c;0,0,0

```
loop_
_space_group_symop_id
_space_group_symop_operation_xyz
1 x,y,z
2 -x,-y,-z

loop_
_atom_site_label
_atom_site_type_symbol
_atom_site_symmetry_multiplicity
_atom_site_Wyckoff_label
_atom_site_fract_x
_atom_site_fract_y
_atom_site_fract_z
_atom_site_occupancy
_atom_site_fract_symmform
Ag1 Ag  2 i 0.01510 0.15646 0.17462  1.00000 Dx,Dy,Dz
Ag2 Ag  2 i 0.50869 0.66189 0.16058  1.00000 Dx,Dy,Dz
Ag3 Ag  1 g 0.00000 0.50000 0.50000  1.00000 0,0,0
Ag4 Ag  1 f 0.50000 0.00000 0.50000  1.00000 0,0,0
F1  F   2 i 0.41040 0.41358 0.41559  1.00000 Dx,Dy,Dz
F2  F   2 i 0.37683 0.09008 0.10663  1.00000 Dx,Dy,Dz
F3  F   2 i 0.12803 0.83808 0.06713  1.00000 Dx,Dy,Dz
F4  F   2 i 0.27912 0.16289 0.73619  1.00000 Dx,Dy,Dz
F5  F   2 i 0.71349 0.51941 -0.06883 1.00000 Dx,Dy,Dz
F6  F   2 i 0.08398 0.49515 0.76612  1.00000 Dx,Dy,Dz
F7  F   2 i 0.84979 0.86031 0.56648  1.00000 Dx,Dy,Dz
F8  F   2 i 0.20863 0.77502 0.39891  1.00000 Dx,Dy,Dz
F9  F   2 i 0.33694 0.77612 0.74518  1.00000 Dx,Dy,Dz
```

**$AgF_4$ *C2/m* @ 60 GPa**
```
data_findsym-output
_audit_creation_method FINDSYM

_cell_length_a    6.2396530000
_cell_length_b    3.8386620000
_cell_length_c    4.4738080000
_cell_angle_alpha 90.0000000000
_cell_angle_beta  133.1517330000
_cell_angle_gamma 90.0000000000
_cell_volume      78.1753375038

_symmetry_space_group_name_H-M "C 1 2/m 1"
_symmetry_Int_Tables_number 12
_space_group.reference_setting '012:-C 2y'
_space_group.transform_Pp_abc a,b,c;0,0,0

loop_
_space_group_symop_id
_space_group_symop_operation_xyz
1 x,y,z
2 -x,y,-z
3 -x,-y,-z
4 x,-y,z
5 x+1/2,y+1/2,z
6 -x+1/2,y+1/2,-z
7 -x+1/2,-y+1/2,-z
8 x+1/2,-y+1/2,z
```

```
loop_
_atom_site_label
_atom_site_type_symbol
_atom_site_symmetry_multiplicity
_atom_site_Wyckoff_label
_atom_site_fract_x
_atom_site_fract_y
_atom_site_fract_z
_atom_site_occupancy
_atom_site_fract_symmform
Ag1  Ag   2 b 0.00000 0.50000 0.00000 1.00000 0,0,0
F1   F    4 i 0.87263 0.00000 0.17192 1.00000 Dx,0,Dz
F2   F    4 i 0.31331 0.00000 0.44820 1.00000 Dx,0,Dz
```